\newcommand{\po}{p_1}
\newcommand{\pt}{p_2}
\newcommand{\pop}{p_1^\prime}
\newcommand{\ptp}{p_2^\prime}
\newcommand{\popp}{p_1^{\prime\prime}}
\newcommand{\ptpp}{p_2^{\prime\prime}}
\newcommand{\pob}{\bar{p}_1}
\newcommand{\ptb}{\bar{p}_2}
\newcommand{\hatq}{\hat{q}}
\newcommand{\smn}{s_-}
\newcommand{\spl}{s_+}
\newcommand{\pole}{{\cal P}_e^z}
\newcommand{\poley}{{\cal P}_e^y}
\newcommand{\polg}{{\cal P}_\gamma}
\newcommand{\apr}{{\cal A}}
\newcommand{\elab}{E^\prime_{\rm lab}}
\newcommand{\klab}{k^\prime_{\rm lab}}
\newcommand{\alr}{A_{LR}}

\newcommand{\etal}{{\it et al.}}

\newcommand{\lsim}{{\mathrel{\mathpalette\@versim<}}}
\newcommand{\gsim}{{\mathrel{\mathpalette\@versim>}}}

\documentstyle[aps,prd,preprint,epsfig]{revtex}
\begin{document}
%\draft
\preprint{\vbox{\hsize=120pt\noindent SLAC--PUB--7701 \\  November 1997}}

\title{A Complete Order-$\alpha^3$ Calculation of the Cross Section for 
Polarized Compton Scattering$^\dagger$}
%% repeat the \author\address pair as needed
\author{Morris L. Swartz}
\address{Stanford Linear Accelerator Center\\ Stanford University, 
Stanford, Ca. 94309\\}
%\date{\today}
\maketitle
\begin{abstract}
The construction of a computer code to calculate the cross sections for the 
spin-polarized processes $e^-\gamma\to e^-\gamma,e^-\gamma\gamma,e^-e^+e^-$ 
to order-$\alpha^3$ is described.  The code calculates cross sections for 
circularly-polarized initial-state photons and arbitrarily polarized 
initial-state electrons.  The application of the code to the SLD Compton 
polarimeter indicates that the order-$\alpha^3$ corrections produce a 
fractional shift in the SLC polarization scale of $-$0.1\% which is too small 
and of the wrong sign to account for the discrepancy in the Z-pole asymmetries 
measured by the SLD Collaboration and the LEP Collaborations. 
\end{abstract}
\vskip 0.3in
\begin{center}
{\rm Submitted to {\em Physical Review D}}
\end{center}
\vskip 1.00in 
\vbox{\footnotesize\renewcommand{\baselinestretch}{1}\noindent
 $^\dagger$This work was supported by Department of Energy
  contract DE-AC03-76SF00515}

% body of paper here

\section{Introduction}
\label{sec:intro}

For some years, polarized Compton scattering, the scattering of 
circularly-polarized photons by spin-polarized electrons, has been used 
to measure the degree of polarization of one particle or the other.  
Circularly-polarized gamma-ray photons from nuclear decays have been 
polarization-analyzed by measuring the asymmetry in the rates of 
backscattering from magnetized iron foils (as the foil magnetization is 
reversed).  Similarly, the polarization of electrons in high-energy storage 
rings and accelerators has been determined from scattering asymmetries of 
the accelerated beam with beams of optical laser photons.  Until quite 
recently, all such measurements have made use of tree-level (order-$\alpha^2$) 
expressions for the polarized Compton scattering cross section.

Part of the reason for this has been the unavailability of a 
next-to-leading-order calculation that is packaged in an easily usable form.  
The first calculation of the order-$\alpha^3$ virtual and real-soft-photon 
corrections to unpolarized Compton scattering was published by Brown and 
Feynman in 1952 \cite{ref:bf}.  This calculation was not confirmed until 1972 
when Tsai, DeRaad, and Milton (TDM) published the same corrections for the 
polarized case \cite{ref:tdrm}.  The TDM calculation, by itself, is sufficient 
to interpret the results of measurements involving longitudinally polarized 
electrons for which the presence of additional energetic photons in the final 
state can be excluded.  This is often the case for measurements of gamma-rays 
that have been scattered from magnetized iron targets.  However, 
accelerator-based polarimeters are often designed to measure transverse 
electron polarization and generally cannot distinguish between single-photon 
and multiple-photon final states.  These shortcomings were addressed in 1987 
by G\'ongora and Stuart (GS) who published the matrix elements for the 
hard-photon corrections in a spinor-product form that is suitable for 
numerical evaluation \cite{ref:gs}.  Their publication also includes 
spinor-product expressions for the matrix elements of six gauge-invariant 
tensors used by TDM to calculate their result.  These expressions permit the 
application of the TDM virtual corrections to the case of general initial- 
and final-state electron spin directions.  Finally, in 1989, the complete set 
of virtual, soft-photon, and hard-photon radiative corrections to polarized 
Compton scattering was calculated independently by Veltman \cite{ref:veltman}.  
Veltman's paper describes her calculation qualitatively and presents a result 
in numerical form for two specific cases of an accelerator-based longitudinal 
polarimeter.   Unfortunately, it does not include detailed expressions for the 
final result nor is the result checked against the TDM or GS calculations.

One of the specific cases discussed by Veltman, the case of a 50~GeV 
longitudinally-polarized electron colliding with a 2.34~eV photon, is quite 
close to that of the SLD Compton polarimeter (a 45.65~GeV 
longitudinally-polarized electron colliding with a 2.33~eV photon).  This 
polarimeter is a key component in the measurement of the left-right $Z$-boson 
production asymmetry $A_{LR}^0$ which has been performed over several years by 
the SLD Collaboration \cite{ref:alr}.  At the current time, the measured value 
of $A_{LR}^0$ is approximately 8\% larger than the value of the comparable 
quantity extracted from measurements of six different $Z$-pole asymmetries by 
the four LEP Collaborations \cite{ref:LEP}.  Since the SLD and LEP measurements 
differ by approximately three standard deviations, the discrepancy is more 
likely to be due to systematic effects than to statistical fluctuations.  One 
possible systematic effect is the absence of radiative corrections from the 
interpretation of the SLD polarimeter data.  Veltman's calculation implies that 
radiative corrections would shift the SLD polarization measurements by $-$0.1\% 
of themselves which is far too small and of the wrong sign to account for the 
8\% discrepancy.

This paper describes a complete order-$\alpha^3$ calculation of polarized 
Compton Scattering.  It was undertaken primarily to check the calculation of 
Veltman and to determine if radiative corrections to Compton scattering could 
be responsible for discrepancy between the LEP and SLD measurements of $Z$-pole 
asymmetries.  A second goal was to develop a computer code which could applied 
to a variety of present and future experimental situations.  The main 
ingredients of this code, the TDM and GS calculations, are sufficient for all 
present day experimental situations.  However, it is likely that a very high 
energy linear electron-positron collider will be constructed somewhere in the 
world in the coming decade.  Polarized beams are planned for all of the designs 
now under discussion.   All of these projects incorporate Compton Scattering 
polarimeters into the optical designs of their final focusing systems.  If 
these polarimeters use optical lasers, the $e^-\gamma$ center-of-mass energies 
will be above threshold for the production of final state $e^+e^-$ pairs.  
Since the process $e^-\gamma\to e^-e^+e^-$ occurs at order-$\alpha^3$, it must 
also be included in the computer code.  A calculation of the matrix element for 
this process, based upon the techniques of Ref.~\cite{ref:gs}, is described in 
Section~\ref{subsec:threee}.

The following sections of this paper describe the construction and operation of 
the Fortran-code COMRAD which calculates the order-$\alpha^3$ cross section for 
polarized Compton Scattering.  Section~\ref{sec:ingred} describes the 
ingredients of the calculations which the code is based. 
Section~\ref{sec:implement} describes the actual implementation of the various 
calculations and several cross checks that were performed.  
Section~\ref{sec:results} describes the application of the code to several 
cases of interest.  And finally, Section~\ref{sec:summary} summarizes the 
preceding sections.

\section{Ingredients}
\label{sec:ingred}

This section describes the ingredients used to construct the code COMRAD.  The 
hard photon corrections, virtual photon corrections, soft photon corrections, 
and $e^-e^+e^-$ cross sections are discussed in the following sections.  Since 
the $e^-e^+e^-$ cross section calculation makes use of the techniques used to 
calculate the hard photon corrections, some technical details are presented in 
Section~\ref{subsec:HPC} that facilitate the description of the original work 
presented in Section~\ref{subsec:threee}.

\subsection{Hard-Photon Corrections}
\label{subsec:HPC}

The calculation of the cross section for the process $e^-\gamma\to 
e^-\gamma\gamma$ is based upon the matrix element calculation of G\'ongora 
and Stuart \cite{ref:gs}.  Their calculation is the first application of 
numerical spinor product techniques \cite{ref:ks} 
to a case involving massive spinors.  These techniques 
allow one to express any amplitude as a function of the scalar products of two 
massless spinors $u_\pm(p)$ and their conjugates $\bar{u}_\pm(p)$.  The 
subscripts refer to positive and negative helicity states of a massless fermion 
of momentum $p$.  The only two non-vanishing scalar products,
\begin{eqnarray}
\spl(p_1,p_2) & = & \bar{u}_+(p_1)u_-(p_2) = -\spl(p_2,p_1) \\
\smn(p_1,p_2) & = & \bar{u}_-(p_1)u_+(p_2) = -\spl(p_1,p_2)^*,
\end{eqnarray}
are easy to evaluate numerically.  G\'ongora and Stuart define the photon 
polarization vector in terms of these quantities so that it is free of 
axial-vector components and can be used with massive currents,
\begin{equation}
\epsilon^\mu_\pm(q,\hat{q}) = 
\pm\frac{1}{\sqrt{2}s_\pm(\hat{q},q)}\bar{u}_\pm(\hat{q}) \gamma^\mu u_\pm(q),
\label{eq:epsdef}
\end{equation}
where: the $\pm$ subscript refers to the helicity of initial-state photons 
(final-state photons have opposite helicities), $q$ is the photon momentum, 
and $\hat{q}$ is an arbitrary massless vector.  Massive spinors of arbitrary 
spin direction are defined in terms of massless spinors as follows,
\begin{eqnarray}
u(p,s) & = & \frac{s_+(p_1,p_2)}{m} u_+(p_1) + u_-(p_2) \\
\bar{u}(p,s) & = & -\frac{s_-(p_1,p_2)}{m} \bar{u}_+(p_1) + \bar{u}_-(p_2)
\label{eq:upsdef}
\end{eqnarray}
where $m$ is the electron mass and the massless vectors, $p_1$ and $p_2$, 
are defined in terms of the momentum and spin vectors, $p$ and $s$, as follows,
\begin{eqnarray}
p_1 & = & \frac{1}{2}\left(p+ms\right) \\
p_2 & = & \frac{1}{2}\left(p-ms\right). 
\label{eq:p1p2def}
\end{eqnarray}

\begin{figure}[t]
\centering
\epsfig{file=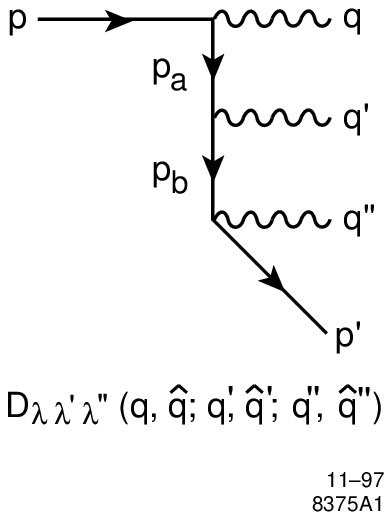,height=2.5in}
\caption{Amplitude for the process $e^-\to e^-\gamma\gamma\gamma$ (time flows 
left to right).}
\label{fg:one}
\end{figure}

The actual calculation involves the evaluation of a single Feynman amplitude 
$D_{\lambda\lambda^\prime\lambda^{\prime\prime}}(q,\hat{q}; 
q^\prime,\hat{q}^\prime; q^{\prime\prime},\hat{q}^{\prime\prime})$ for the 
process $e^-(s)\to 
e^-(s^\prime)\gamma(\lambda)\gamma(\lambda^\prime)
\gamma(\lambda^{\prime\prime})$ (shown in Fig.~\ref{fg:one}) where: $\lambda$, 
$\lambda^\prime$, and $\lambda^{\prime\prime}$ label the helicities of the 
three photons ($+$ or $-$); and $s$ and $s^\prime$ are the spin-vectors of the 
initial- and final-state electrons, respectively.  The matrix element ${\cal 
M}_{\lambda;\lambda^\prime\lambda^{\prime\prime}}(s,s^\prime)$ for the process 
$e^-(s)\gamma(\lambda)\to 
e^-(s^\prime)\gamma(\lambda^\prime)\gamma(\lambda^{\prime\prime})$ can then be 
constructed from the function $D_{\lambda\lambda^\prime\lambda^{\prime\prime}}$ 
by reversing the momenta of single photons and by interchanging the momenta and 
helicities of the remaining identical photons,
\begin{eqnarray}
{\cal M}_{\lambda;\lambda^\prime\lambda^{\prime\prime}}(s,s^\prime) & = & 
D_{\lambda\lambda^\prime\lambda^{\prime\prime}}(-q,\hat{q}; 
q^\prime,\hat{q}^\prime; q^{\prime\prime},\hat{q}^{\prime\prime}) +
D_{\lambda\lambda^{\prime\prime}\lambda^\prime}(-q,\hat{q}; 
q^{\prime\prime},\hat{q}^{\prime\prime}; q^\prime,\hat{q}^\prime) \nonumber  \\ 
& + &
D_{\lambda^\prime\lambda\lambda^{\prime\prime}}(q^\prime,\hat{q}^\prime; 
-q,\hat{q}; q^{\prime\prime},\hat{q}^{\prime\prime}) +
D_{\lambda^{\prime\prime}\lambda\lambda^\prime}(q^{\prime\prime},
\hat{q}^{\prime\prime}; -q,\hat{q}; q^\prime,\hat{q}^\prime) \nonumber \\  
& + &
D_{\lambda^{\prime\prime}\lambda^\prime\lambda}(q^{\prime\prime},
\hat{q}^{\prime\prime}; q^\prime,\hat{q}^\prime; -q,\hat{q}) +
D_{\lambda^\prime\lambda^{\prime\prime}\lambda}(q^\prime,\hat{q}^\prime; 
q^{\prime\prime},\hat{q}^{\prime\prime}; -q,\hat{q}), 
\label{eq:dlll}
\end{eqnarray}
where $q$, $q^\prime$, and $q^{\prime\prime}$ are the momenta of the incident 
and final-state photons, respectively.  Note that each of the six terms in 
Eq.~\ref{eq:dlll} corresponds to an ordinary Feynman diagram.

Two technical issues are relevant to the present discussion and to the 
presentation of Section~\ref{subsec:threee}.  The first issue concerns the 
choice of the arbitrary massless momenta: $\hat{q}$, $\hat{q}^\prime$, and 
$\hat{q}^{\prime\prime}$.  G\'ongora and Stuart point out that one can 
substantially simplify some of the expressions by a judicious choice of these 
auxiliary momenta.  They present results for two equivalently-simple sets of 
momenta.  This approach provides an important cross check (one must find 
identical results for both sets of auxiliary momenta) and greatly facilitated 
the debugging of the GS manuscript and the computer code.  A number of 
typographical errors were discovered in Ref.~\cite{ref:gs} and are listed in 
Appendix~\ref{sec:errata}.  
One should note that the first set of auxiliary momenta (used to 
calculate GS Eqs.~3.3.1-3.10.2) always produces singularities when the initial 
state electron is longitudinally polarized whereas the second set (used to 
calculate GS Eqs.~C.1.1-C.8.2) never develops singularities so long as both 
photons have non-zero energy.

The second issue concerns the evaluation of spinors of negative momenta.  In 
order to preserve the following (very useful) relationship, 
\begin{equation}
\frac{1}{2}\left(1\pm\gamma_5\right)\not\!p = u_\pm(p)\bar{u}_\pm(p),
\end{equation}
it is necessary to define negative momentum spinors in the following manner,
\begin{equation}
\bar{u}_\pm(-p) = i\bar{u}_\pm(p),\quad\quad u_\pm(-p) = iu_\pm(p).
\label{eq:negp}
\end{equation}
This, in turn, implies that spinor products of negative arguments behave as 
follows,
\begin{eqnarray}
s_\pm(-q_1,q_2) & = & s_\pm(q_1,-q_2) = is_\pm(q_1,q_2) \\
s_\pm(-q_1,-q_2) & = & -s_\pm(q_1,q_2),
\end{eqnarray}
and that external photon polarization vectors are invariant under the 
transformation $q\to-q$ (see Eq.~\ref{eq:epsdef}),
\begin{equation}
\epsilon^\mu(-q,\hatq) = \epsilon^\mu(q,\hatq).
\label{eq:epsneg}
\end{equation}

The actual cross section for the process $e^-(s)\gamma(\lambda)\to 
e^-\gamma\gamma$ is calculated in the center-of-mass (cm) frame from the matrix 
element given in Eq.~\ref{eq:dlll} using the following expression,
\begin{equation}
\frac{d^5\sigma^{(1)}_{e\gamma\gamma}}{dE_e^\prime d\Omega_e^\prime 
dE_\gamma^\prime d\phi_\gamma^\prime}(s,\lambda) = 
\frac{1}{64(2\pi)^5E_\gamma(E_e+P_e)} 
\sum_{\lambda^\prime,\lambda^{\prime\prime},s^\prime} \left| {\cal 
M}_{\lambda;\lambda^\prime\lambda^{\prime\prime}}(s,s^\prime) \right|^2,
\label{eq:eggxs}
\end{equation}
where: $E_e$ and $P_e$ are the energy and 3-momentum of the incident electron, 
$E_\gamma$ is the energy of the incident photon, $E_e^\prime$ and 
$\Omega_e^\prime$ are the energy and direction of the final state electron, 
$E_\gamma^\prime$ is then energy of one of the final state photons, and 
$\phi_\gamma^\prime$ is the azimuth of the final state photon with respect to 
the final-state electron direction \cite{ref:phinote}.  Note that 
Eq.~\ref{eq:eggxs} includes a factor of $1/2$ to account for the identical 
photons in the final state.
 
\subsection{Virtual Corrections}
\label{subsec:VC}

The matrix element for the process $e^-(s)\gamma(\lambda)\to 
e^-(s^\prime)\gamma(\lambda^\prime)$ is expressed by Tsai, DeRaad, and Milton 
in the following form \cite{ref:normnote},
\begin{equation}
{\cal M}^{(j)}_{\lambda;\lambda^\prime}(s,s^\prime) = \frac{1}{2m^4}\sum_{i=1}^6
\bar{u}(p^\prime,s^\prime)\epsilon_{\lambda^\prime}(q^\prime)\cdot{\cal L}_i\cdot\epsilon_{\lambda}(q)u(p,s)\ M^{(j)}_i
\label{eq:mattwo}
\end{equation}
where $j=0,1$ labels the order of the matrix element and the six 
gauge-invariant, singularity-free, kinematic-zero-free, Dirac tensors ${\cal 
L}_i$ are defined by Bardeen and Tung \cite{ref:tb}.  The authors calculate the 
six invariant matrix elements $M^{(j)}_i$ within the framework of Schwinger 
source theory to order-$\alpha$ ($M_i^{(0)}$) and to order-$\alpha^2$ 
($M_i^{(1)}$).  They explicitly consider the case that the initial- and 
final-state electrons are longitudinally polarized and express the matrix 
element as a set of six helicity amplitudes (due to the charge-conjugation and 
time-reversal symmetries, only six of the eight matrix elements defined in 
Eq.~\ref{eq:mattwo} are independent) which are linear combinations of the six 
invariant matrix elements.  The helicity amplitudes are then used to derive an 
order-$\alpha^3$ expression for the unpolarized cross section which is found to 
agree with the calculation of Brown and Feynman.  This cross check was found to 
be useful in locating three typographical sign errors in the rendering of the 
helicity amplitudes (which, given the complexity of the expressions, is a 
remarkably small number).  The errors are listed in Appendix~\ref{sec:errata}.

As was mentioned in the introduction, G\'ongora and Stuart supply spinor-product 
expressions for the six tensors,
\begin{equation}
T^i_{\lambda\lambda^\prime}(s,s^\prime)=\bar{u}(p^\prime,s^\prime)
\epsilon_{\lambda^\prime}(q^\prime)\cdot{\cal 
L}_i\cdot\epsilon_{\lambda}(q)u(p,s),
\end{equation}
which permits the application of the virtual corrections contained in the 
invariant matrix elements $M_i^{(1)}$ to the case of general initial-state and 
final-state electron spin directions.  To make use of these, the system of six 
equations which define the helicity amplitudes was inverted to extract the 
$M_i$.

The order-$\alpha^2$ and order-$\alpha^3$ cross sections for the process 
$e^-(s)\gamma(\lambda)\to e^-\gamma$ are then calculated (in the cm-frame) from 
the order-$\alpha$ and order-$\alpha^2$ matrix elements as follows,
\begin{eqnarray}
\frac{d^2\sigma^{(0)}_{e\gamma}}{d\Omega_e^\prime}(s,\lambda) & = & 
\frac{1}{64\pi^2\left[m^2+2E_\gamma(E_e+P_e)\right]} 
\sum_{\lambda^\prime,s^\prime} \left|{\cal 
M}_{\lambda;\lambda^\prime}^{(0)}(s,s^\prime)\right|^2 \label{eq:xsegz}\\
\frac{d^2\sigma^{(1V)}_{e\gamma}}{d\Omega_e^\prime}(s,\lambda) & = & 
\frac{1}{64\pi^2\left[m^2+2E_\gamma(E_e+P_e)\right]} 
\sum_{\lambda^\prime,s^\prime} 2\Re\left[{\cal 
M}_{\lambda;\lambda^\prime}^{(0)}(s,s^\prime){\cal 
M}_{\lambda;\lambda^\prime}^{(1)*}(s,s^\prime)\right],
\label{eq:xsegov}
\end{eqnarray}
where the order-$\alpha^3$ cross section has been labeled as 
$\sigma^{(1V)}_{e\gamma}$ to explicitly indicate that Eq.~\ref{eq:xsegov} 
describes virtual corrections only.

\subsection{Soft-Photon Corrections}
\label{subsec:SPC}

The order-$\alpha^3$ cross section defined in Eq.~\ref{eq:xsegov} contains a 
term that depends logarithmically upon a small, but nonzero, fictitious photon 
mass ($m_\gamma$) used to regulate singularities in the virtual corrections.  
This unphysical term is cancelled by a similar term which arises in the cross 
section for $e^-\gamma\to e^-\gamma\gamma$ for slightly massive photons.  Brown 
and Feynman discuss this point at some length in Ref.~\cite{ref:bf}.  Since 
events with additional photons of energy less than some small value 
$k_\gamma^{min}$ are experimentally indistinguishable from the two-body final 
state, they explicitly integrate the three-body cross section over the extra 
photon momenta $q^{\prime\prime}$ in the region 
$m_\gamma<E_\gamma^{\prime\prime}<k_\gamma^{min}$.  The resulting soft-photon 
cross section is approximately equal to the product of a function $J$ and the 
order-$\alpha^2$ cross section.  The order-$\alpha^3$ 2-body cross section can 
now be defined as a function of $k_\gamma^{min}$,
\begin{eqnarray}
\frac{d^2\sigma^{(1)}_{e\gamma}}{d\Omega_e^\prime}(s,\lambda;k_\gamma^{min}) & 
= & \frac{d^2\sigma^{(1V)}_{e\gamma}}{d\Omega_e^\prime}(s,\lambda;m_\gamma) + 
\int_{m_\gamma}^{k_\gamma^{min}} d^3q^{\prime\prime} 
\frac{d^5\sigma^{(1)}_{e\gamma\gamma}}{d\Omega_e^\prime 
d^3q^{\prime\prime}}(s,\lambda) \nonumber \\
& \simeq & \frac{d^2\sigma^{(1V)}_{e\gamma}}{d\Omega_e^\prime}(s,\lambda;m_\gamma) 
+ J(m_\gamma,k_\gamma^{min},\Omega_e^\prime) 
\frac{d^2\sigma^{(0)}_{e\gamma}}{d\Omega_e^\prime}(s,\lambda),
\label{eq:jdef}
\end{eqnarray}
where $\sigma^{(1)}_{e\gamma}$ is independent of $m_\gamma$.  One should note 
that although Eq.~\ref{eq:jdef} is independent of reference frame, the actual 
integration \cite{ref:bf,ref:tdrm} was performed in the rest frame of the 
initial-state electron.  The use of the resulting expression for $J$ implies 
that the quantity $k_\gamma^{min}$ is defined in that frame.

\subsection{The $e^-e^+e^-$ Final State}
\label{subsec:threee}

The cross section for the process $e^-(s)\gamma(\lambda)\to 
e^-(s^\prime)e^+(\bar{s})e^-(s^{\prime\prime})$ is calculated using the massive 
spinor-product techniques given in Ref.~\cite{ref:gs}.  Since the work 
described there doesn't involve positrons, its authors did not define massive 
positron spinors.  It is extremely straightforward to do this from 
Eqs.~\ref{eq:upsdef} and \ref{eq:p1p2def} by making the replacement $m\to -m$.  
This interchanges the massless momentum vectors, $p_1\leftrightarrow p_2$, and 
yields the following massive positron spinors,   
\begin{eqnarray}
v(p,s) & = & \frac{s_+(p_1,p_2)}{m} u_+(p_2) + u_-(p_1) \\
\bar{v}(p,s) & = & -\frac{s_-(p_1,p_2)}{m} \bar{u}_+(p_2) + \bar{u}_-(p_1).
\label{eq:vpsdef}
\end{eqnarray}
These spinors have the correct normalization and orthogonality properties,
\begin{eqnarray*}
\bar{u}(p,s)u(p,s) & = & 2m, \quad\bar{v}(p,s)v(p,s) = -2m, \\
\bar{u}(p,s)v(p,s) & = & \bar{v}(p,s)u(p,s) = 0.
\end{eqnarray*}
Similarly, the evaluation of a massive spinor with a negative momentum leads to 
the replacements $\{p_1,p_2\}\to\{-p_2,-p_1\}$ and using Eq.~\ref{eq:negp} one 
finds the correct behavior to within extra phases, 
\begin{eqnarray}
v(-p,s) & = & iu(p,s),\quad\bar{v}(-p,s) = i\bar{u}(p,s), \\
u(-p,s) & = & iv(p,s),\quad\bar{u}(-p,s) = i\bar{v}(p,s).
\end{eqnarray}
These extra phases do not occur in the case of external photons (see 
Eq.~\ref{eq:epsneg}) and must be treated with some care.  To avoid disturbing 
the phase relationships between diagrams, a calculation must be formulated so 
that all diagrams contain the same number of momentum-reversed massive spinors.

The actual calculation was carried out by calculating spinor-product 
expressions for two Feynman amplitudes, $D_{1\lambda}$ and $D_{2\lambda}$, for 
the process $\gamma\to e^-e^+e^-e^+$ as shown in Fig.~\ref{fg:two}.  These 
expressions are listed in Appendix~\ref{sec:mateee}.  
The matrix element is the sum of the 
eight diagrams generated by reversing one of the positron momenta and by 
interchanging final state electron momenta (according to Fermi-Dirac 
statistics),
\begin{eqnarray}
{\cal M}_{\lambda}(s,\bar{s},s^\prime,s^{\prime\prime}) & = &
D_{1\lambda}(-p,s;\bar{p},\bar{s};p^\prime,s^\prime;p^{\prime\prime},
s^{\prime\prime}) - D_{1\lambda}(-p,s;\bar{p},\bar{s};p^{\prime\prime},
s^{\prime\prime};p^\prime,s^\prime) \nonumber \\ & - &
D_{1\lambda}(\bar{p},\bar{s};-p,s;p^\prime,s^\prime;p^{\prime\prime},
s^{\prime\prime}) +
D_{1\lambda}(\bar{p},\bar{s};-p,s;p^{\prime\prime},s^{\prime\prime};
p^\prime,s^\prime) \nonumber \\ & + &
D_{2\lambda}(-p,s;\bar{p},\bar{s};p^\prime,s^\prime;p^{\prime\prime},
s^{\prime\prime}) - D_{2\lambda}(-p,s;\bar{p},\bar{s};p^{\prime\prime},
s^{\prime\prime};p^\prime,s^\prime) \nonumber \\ & - &
D_{2\lambda}(\bar{p},\bar{s};-p,s;p^\prime,s^\prime;p^{\prime\prime},
s^{\prime\prime}) +
D_{2\lambda}(\bar{p},\bar{s};-p,s;p^{\prime\prime},s^{\prime\prime};
p^\prime,s^\prime), \label{eq:meee}
\end{eqnarray}
where: $p$ and $s$ are the momentum and spin of the initial-state electron, 
$\bar{p}$ and $\bar{s}$ are the momentum and spin of the final-state positron, 
$p^\prime$ and $s^\prime$ are the momentum and spin of one final-state 
electron, and $p^{\prime\prime}$ and $s^{\prime\prime}$ are the momentum and 
spin of the other final-state electron.
This particular formulation does not benefit from algebraic simplifications due 
to a clever choice of the photon auxiliary momentum $\hat{q}$.  It is possible 
to reduce the total number of terms in the matrix element from 112 to 96 by 
defining four amplitudes instead of two and by choosing $\hat{q}$ 
appropriately.  As formulated, the choice of $\hat{q}$ is truly arbitrary.

The matrix element given in Eq.~\ref{eq:meee} is converted into a cross section 
with an expression that is very similar to Eq.~\ref{eq:eggxs}, 
\begin{equation}
\frac{d^5\sigma^{(1)}_{eee}}{dE_e^\prime d\Omega_e^\prime dE_e^{\prime\prime} 
d\phi_e^{\prime\prime}}(s,\lambda) = \frac{1}{64(2\pi)^5E_\gamma(E_e+P_e)} 
\sum_{\bar{s},s^\prime,s^{\prime\prime}} \left| {\cal 
M}_{\lambda}(s,\bar{s},s^\prime,s^{\prime\prime}) \right|^2,
\label{eq:eeexs}
\end{equation}
where $E_e^{\prime\prime}$ is the energy of the second electron and 
$\phi_e^{\prime\prime}$ is the azimuth of the second electron with respect to 
the first electron direction \cite{ref:phinote}.

\section{Implementation}
\label{sec:implement}

The Fortran-code COMRAD consists of three weighted Monte Carlo generators: 
COMTN2, COMEGG, and COMEEE.  These perform integrations of the cross sections 
for the $e^-\gamma$, $e^-\gamma\gamma$, and $e^-e^+e^-$ final states, 
respectively.  Operational details are given in Appendix~\ref{sec:opdet}.

\subsection{Weighting Scheme}

Each of the generators produces events that consist of momentum four-vectors 
of the final-state state particles in the laboratory frame.  Each event 
is also accompanied by a vector of four event weights $W_j$.  The event weights 
are defined as follows:  
\begin{eqnarray}
W_1 & = & \frac{1}{2\rho^{(n)}(x)}\left[\frac{d^n\sigma^{(0)}}
{dx^n}(s,-)+\frac{d^n\sigma^{(0)}}{dx^n}(s,+) \right] \label{eq:wgti} \\
W_2 & = & \frac{1}{2\rho^{(n)}(x)}\left[\frac{d^n\sigma^{(0)}}
{dx^n}(s,-)-\frac{d^n\sigma^{(0)}}{dx^n}(s,+) \right] \label{eq:wgtii} \\
W_3 & = & \frac{1}{2\rho^{(n)}(x)}\left[\frac{d^n\sigma^{(1)}}
{dx^n}(s,-)+\frac{d^n\sigma^{(1)}}{dx^n}(s,+) \right] \label{eq:wgtiii} \\
W_4 & = & \frac{1}{2\rho^{(n)}(x)}\left[\frac{d^n\sigma^{(1)}}
{dx^n}(s,-)-\frac{d^n\sigma^{(1)}}{dx^n}(s,+) \right], \label{eq:wgtiv} 
\end{eqnarray}
where $n$ is the dimensionality of the integrated space ($n=2$ for the 
$e^-\gamma$ final state, and $n=5$ for the three-body final states) and 
$\rho^{(n)}$ is the density of trials in that space.  

The sums of the weights yield partly or fully integrated cross sections.  It is 
convenient to define the following notation for these sums,
\begin{eqnarray}
&&\sigma^{(0)}_u(x^\prime) = \sum_i W_1^i \\
&&\sigma^{(0)}_p(s;x^\prime) = \sum_i W_2^i \\
&&\sigma^{(1)}_u(x^\prime) = \sum_i W_3^i \\
&&\sigma^{(1)}_p(s;x^\prime) = \sum_i W_3^i, 
\end{eqnarray}
where the variables $x^\prime$ define the kinematical binning chosen for a 
particular problem.  The sums of the $W_1$ and $W_3$ weights yield the 
order-$\alpha^2$ and order-$\alpha^3$ unpolarized cross sections, 
respectively.  The sums of the $W_2$ and $W_4$ weights yield the 
order-$\alpha^2$ and order-$\alpha^3$ polarized cross sections and depend upon 
the initial spin direction $s$ (all cross sections are given in millibarns).  
It is also convenient to define notation for the fully corrected cross sections 
and for the asymmetry functions,
\begin{eqnarray}
&&\sigma_u(x^\prime) = \sigma^{(0)}_u(x^\prime)+\sigma^{(1)}_u(x^\prime) \\
&&\sigma_p(s;x^\prime) = \sigma^{(0)}_p(s;x^\prime)+\sigma^{(1)}_p(s;x^\prime) \\
&&A^{(0)}(s;x^\prime) = \frac{\sigma^{(0)}_p(s;x^\prime)} 
{\sigma^{(0)}_u(x^\prime)} \\
&&A(s;x^\prime) = \frac{\sigma_p(s;x^\prime)} {\sigma_u(x^\prime)}. 
\end{eqnarray}

Note that the polarized cross sections are chosen to be the differences of the 
negative-helicity photon cross sections and the positive-helicity photon cross 
sections.  In particle physics terminology, these are called 
left-handed-helicity and right-handed-helicity photons, respectively.  One 
should note that in optics terminology, a negative-helicity photon is called 
Right-Circularly-Polarized (RCP) and a positive-helicity photon is called 
Left-Circularly-Polarized (LCP).

The generation of multiple weights per event trial allows the user to 
significantly improve the statistical power of a given set of Monte Carlo 
trials.  The uncertainty on any function of the four quantities $\sigma_j$ = 
\{$\sigma_u^{(0)}$, $\sigma_p^{(0)}$, $\sigma_u^{(1)}$, $\sigma_p^{(1)}$\} is 
always smaller when generated with correlated weights than separate, 
uncorrelated calculations would yield.  The correct estimate of the statistical 
uncertainty on any such function requires that the user accumulate the full 
4$\times$4 error matrix $E_{jk}$,
\begin{equation}
E_{jk}=\sum_{i=1}^{\rm ntrial} W^i_jW^i_k,
\end{equation}
where the sum is over all event trials.   This matrix must then be propagated 
correctly to the final result.  As an example, consider the calculation of the 
uncertainty on the quantity, $\Delta A$, which is the difference of the full, 
order-$\alpha^3$-corrected polarized asymmetry and the order-$\alpha^2$ 
asymmetry, 
\begin{equation}
\Delta A(s;x^\prime) = A(s;x^\prime) - A^{(0)}(s;x^\prime) = 
\frac{\sigma_p^{(0)}+\sigma_p^{(1)}}{\sigma_u^{(0)}+\sigma_u^{(1)}} - 
\frac{\sigma_p^{(0)}}{\sigma_u^{(0)}} \label{eq:dadef}
\end{equation}
The correct uncertainty on this quantity is given by the expression,
\begin{equation}
\delta(\Delta A) = \sum_{j,k=1}^4 \frac{\partial\Delta A}{\partial\sigma_j} 
E_{jk} \frac{\partial\Delta A}{\partial\sigma_k}
\end{equation}
where $j$ and $k$ label the four cross sections.

\subsection{Cross Checks}

The code COMRAD has been checked in a number of ways.  The order-$\alpha^3$ 
unpolarized cross section $\sigma^{(1)}_{e\gamma}$ calculated from the 
unpolarized initial-state by COMTN2 is numerically identical to the one 
calculated from the diagnostic expression given by TDM in Ref.~\cite{ref:tdrm} 
and to one given by Brown and Feynman in Ref.~\cite{ref:bf}.  It is verified 
that this cross section is rigorously independent of the value chosen for the 
photon mass $m_\gamma$.  It is also verified that the order-$\alpha^3$ 
polarized cross section is invariant under helicity-flips of both incident 
particles (as required by parity invariance).

\begin{figure}[t]
\centering
\epsfig{file=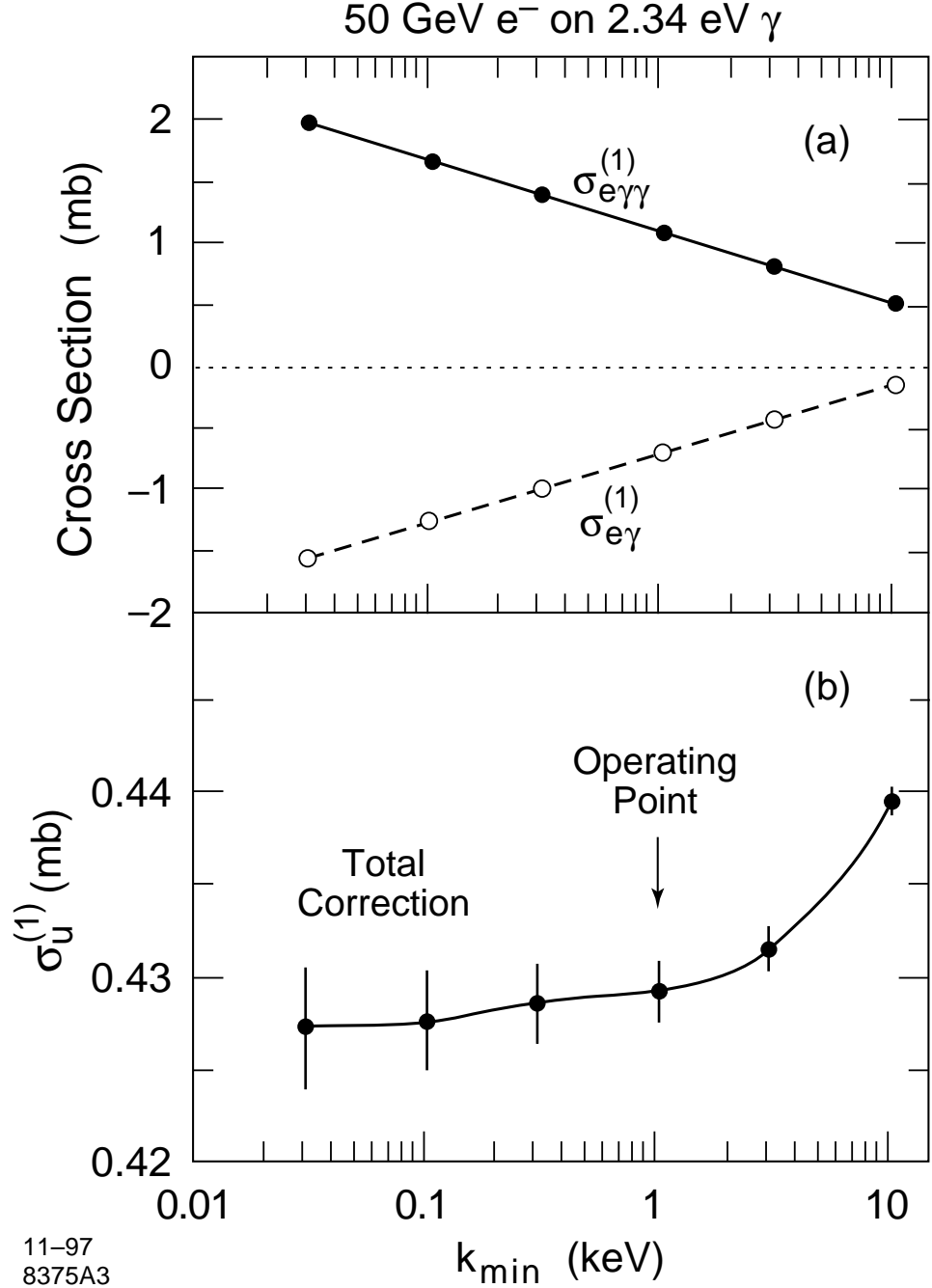,height=7in}
\caption{The unpolarized virtual plus soft-photon cross section 
$\sigma^{(1)}_{e\gamma}$ and hard-photon cross section 
$\sigma^{(1)}_{e\gamma\gamma}$ are shown in part~(a) as functions of 
$k_\gamma^{min}$ for the case of a 50~GeV electron colliding with a 2.34~eV 
photon.  The sum of these cross sections $\sigma_u^{(1)}$ is shown in 
part~(b).}
\label{fg:three}
\end{figure}

The hard-photon cross section calculated by COMEGG is verified to be 
independent of the choice of photon auxiliary momenta.  The polarized 
hard-photon cross section is found to be invariant under helicity-flips of both 
incident particles.  The dependence of the cross sections 
$\sigma^{(1)}_{e\gamma}$ and $\sigma^{(1)}_{e\gamma\gamma}$ on $k_\gamma^{min}$ 
is shown in part~(a) of Fig.~\ref{fg:three} for the case of a 50~GeV electron 
colliding with a 2.34~GeV photon (one of the cases considered by Veltman in 
Ref.~\cite{ref:veltman}).  Note that the cross-sections vary by approximately 
1.4~mb as $k_\gamma^{min}$ is varied from 30~eV to 10~KeV.  The sum of the 
cross sections, $\sigma^{(1)}_u$, is shown in part~(b) of the figure 
and is constant at 0.002~mb level until $k_\gamma^{min}$ reaches several 
percent of the maximum photon energy and the two-body approximation for the 
soft-photon cross section begins to fail.  Even then, the 10~KeV point differs 
by only 0.012~mb from the 30~eV point.

The $e^-e^+e^-$ cross section calculated by COMEEE is found to be independent 
of the choice of photon auxiliary momentum.  The polarized $e^-e^+e^-$ cross 
section is found to be invariant under helicity-flips of both incident 
particles.

\begin{figure}[t]
\centering
\epsfig{file=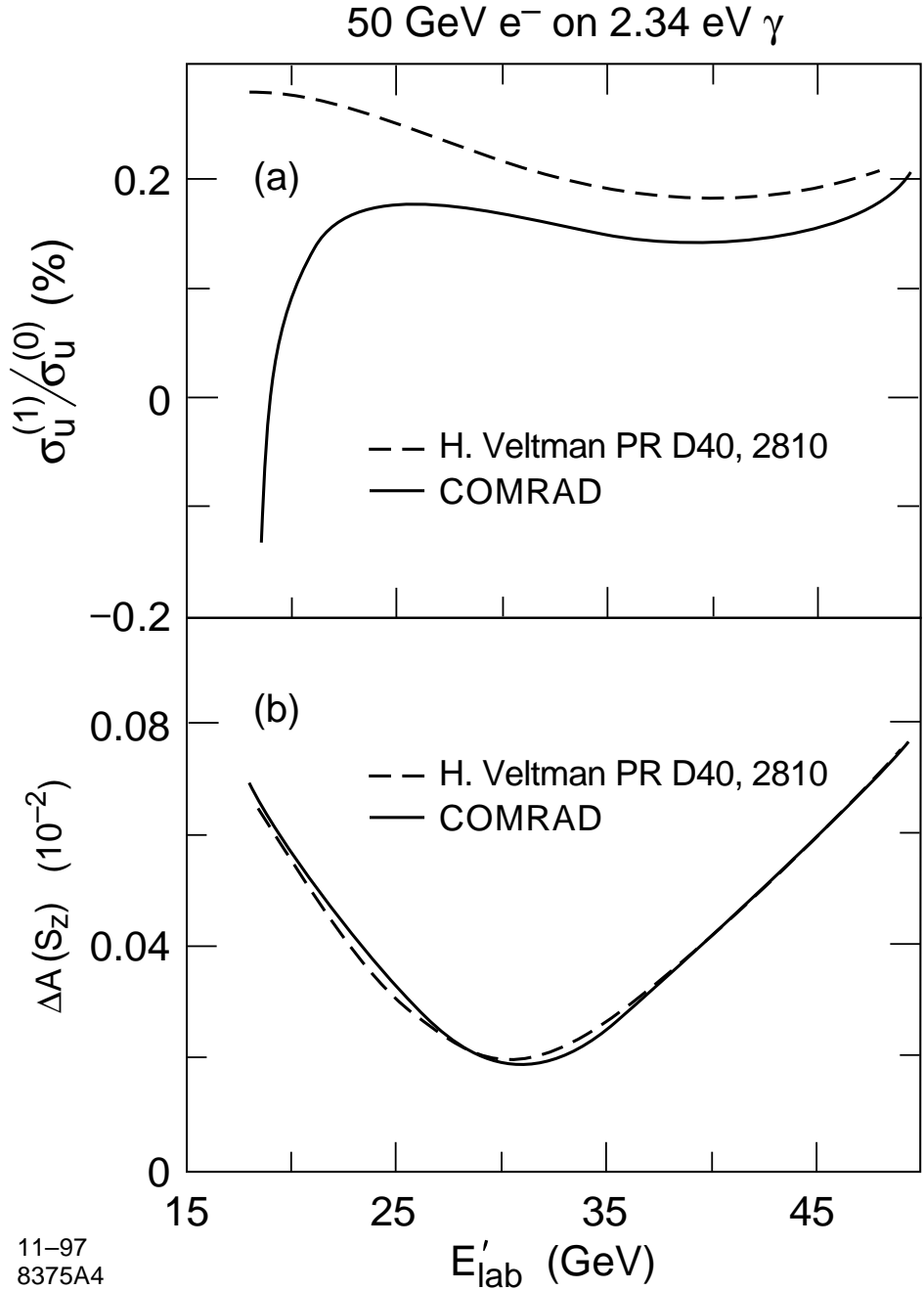,height=7in}
\caption{The ratio $\sigma^{(1)}_u/\sigma^{(0)}_u$ is shown in part~(a) 
as a function of the laboratory energy of the scattered electron for the case 
of a 50~GeV electron colliding with a 2.34~eV photon.  The difference in the 
fully order-$\alpha^3$ corrected asymmetry and the order-$\alpha^2$ asymmetry 
is shown as a function of the laboratory energy of the scattered electron in 
part~(b).  The COMRAD calculation is shown as solid lines and the 
calculation of Ref.~\protect\cite{ref:veltman} is shown as dashed lines.}
\label{fg:four}
\end{figure}

The sum of the virtual, soft-photon, and hard-photon cross sections calculated 
by COMRAD is compared with the numerical result presented in 
Ref.~\cite{ref:veltman} for the case of a 50~GeV electron colliding with a 
2.34~GeV photon.  The ratio of the unpolarized cross sections 
$\sigma^{(1)}_u/\sigma^{(0)}_u$ is presented as a function of the laboratory 
energy of the scattered electron $E^\prime_{lab}$ in part~(a) of 
Fig.~\ref{fg:four}.  The COMRAD calculation predicts that 
$\sigma^{(1)}_u/\sigma^{(0)}_u$ increases from from $-$0.14\% near the 
kinematical edge at 17.90~GeV to $+$0.2\% near the beam energy.  The Veltman 
calculation predicts that the ratio decreases from $+$0.3\% near the edge to 
$+$0.2\% near the beam energy.  The physically correct behavior follows from a 
simple kinematical analysis.  In the center-of-mass frame, the emission of an 
additional photon reduces the energy and momentum available to the scattered 
electron.  Given the large mass of the electron, the fractional change in the 
momentum $P^{\prime}_e$ is larger than than fractional change in the energy 
$E^\prime_e$.  The laboratory energy of a backscattered electron is given by 
the following expression,
\begin{equation}
E^\prime_{lab} = \gamma\left(E^\prime_e-P^{\prime}_e\right),
\end{equation}
where $\gamma$ is the Lorentz factor for the highly-boosted cm-frame (the 
velocity is assumed to be one).  It is straightforward to show that although 
$E^\prime_e$ and $P^\prime_e$ are decreased by the emission of an additional 
photon, the difference $E^\prime_e-P^\prime_e$ increases.  The laboratory 
energy of the backscattered electron is therefore {\it increased} by the 
emission of an additional photon.  It is clear that photon emission depopulates 
the Compton kinematical edge region and that $\sigma^{(1)}_u$ should be 
negative near the endpoint.

The order-$\alpha$ correction to the longitudinal polarization asymmetry is 
shown in part~(b) of Fig.~\ref{fg:four}.  The quantity $\Delta 
A(s_z;\elab)$ (defined in Eq.~\ref{eq:dadef}) predicted by COMRAD is compared 
with the similar quantity given in Ref.~\cite{ref:veltman}.  Good agreement is 
observed. 

\section{Results}
\label{sec:results}

This section describes the application of the COMRAD code to several 
accelerator-based polarimetry cases.  The first case (the SLD polarimeter) 
deals with the detection of the scattered electrons to measure longitudinal 
polarization.  The second case (the HERA polarimeters) involves the detection 
of scattered photons to measure longitudinal and transverse electron 
polarization.  The final case (a Linear Collider polarimeter) illustrates the 
detection of final state electrons when there is sufficient energy to produce 
the $e^-e^+e^-$ final state.

\subsection{The SLD Polarimeter} 

The SLD Polarimeter \cite{ref:alr} is located 33~m downstream of the SLC 
interaction point (IP).  After the 45.65~GeV longitudinally-polarized electron 
beam passes through the IP and before it is deflected by dipole magnets,
it collides with a 2.33~eV circularly-polarized photon beam produced by a 
pulsed frequency-doubled Nd:YAG laser.  The scattered and unscattered 
components of the electron beam are separated by a dipole-quadrupole 
spectrometer.  The scattered electrons are dispersed horizontally and exit the 
vacuum system through a thin window.  A multichannel Cherenkov detector 
observes the scattered electrons in the interval from 17 to 27~GeV/c.  

\begin{figure}[t]
\centering
\epsfig{file=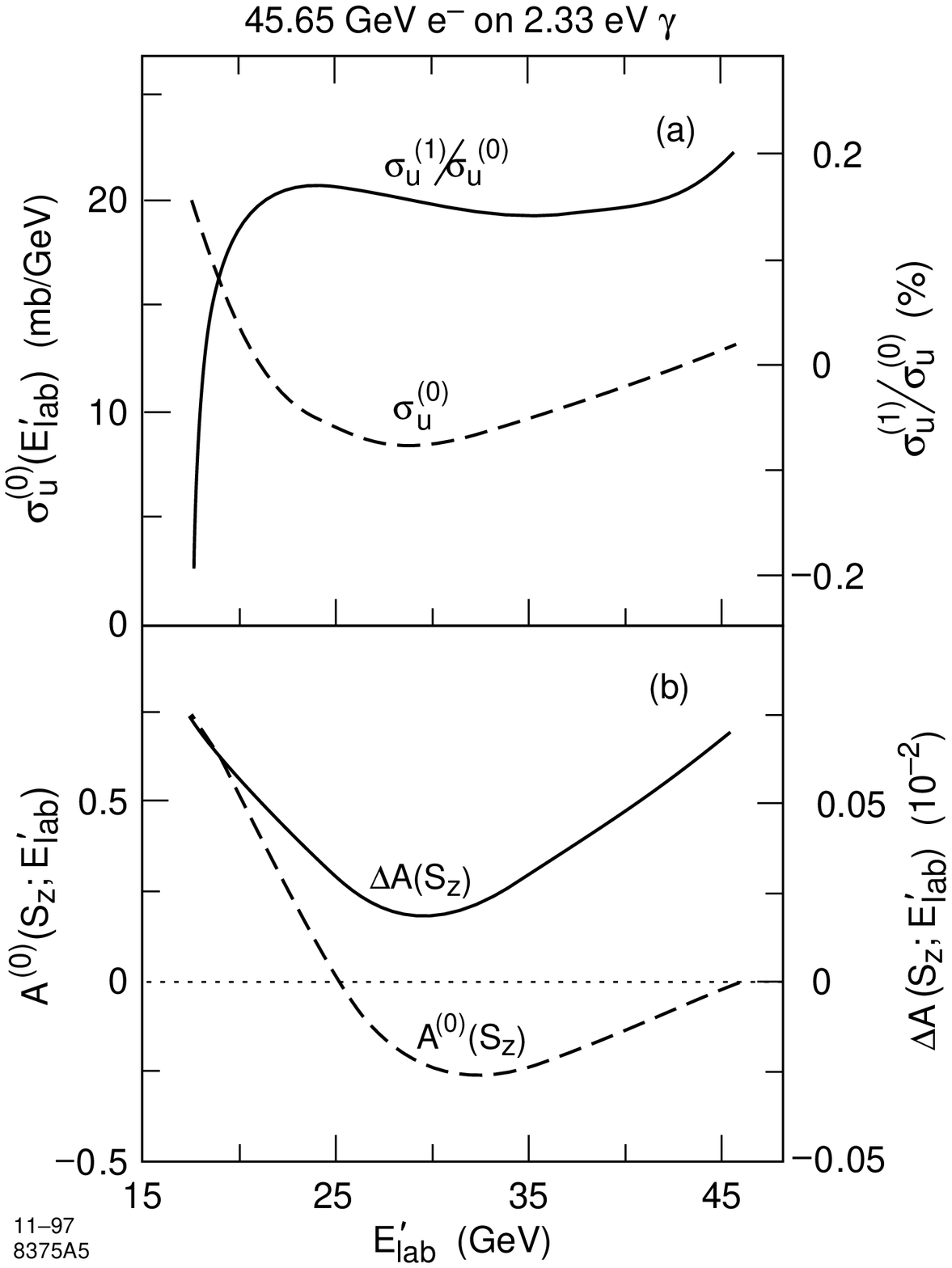,height=7in}
\caption{The order-$\alpha^2$ cross section $\sigma^{(0)}_u$ and the ratio 
$\sigma^{(1)}_u/\sigma^{(0)}_u$ are shown in part~(a) as functions of the 
laboratory energy of the scattered electron for the case of a 45.65~GeV 
electron colliding with a 2.33~eV photon.  The order-$\alpha^2$ asymmetry 
function and the difference in the fully order-$\alpha^3$ corrected asymmetry 
and the order-$\alpha^2$ asymmetry are shown as functions of the laboratory 
energy of the scattered electron in part~(b).}
\label{fg:five}
\end{figure}

The helicities of the electron and photon beams are changed on each beam pulse 
according to pseudo-random sequences.  Each channel of the Cherenkov detector 
measures the asymmetry in the signals $S_j$ observed when the electron and 
photon spins are parallel ($|J_Z|=3/2$) and anti-parallel ($|J_Z|=1/2$),
\begin{equation}
A^C_j=\frac{S_j(3/2) - S_j(1/2)}{S_j(3/2) + S_j(1/2)}=\pole\polg\apr_j
\end{equation}
where: $j$ labels the channels of the detector, $\pole$ is the electron beam 
polarization, $\polg$ is the photon polarization, and $\apr_j$ is the analyzing 
power of the $j^{th}$ channel.  The analyzing powers are defined in terms of 
the Compton scattering cross section and the response function of each channel 
$R_j$,
\begin{equation}
\apr_j = \frac{\int d\elab \sigma_u(\elab)R_j(\elab)A(s_z;\elab)} 
{\int d\elab \sigma_u(\elab)R_j(\elab)} ,
\end{equation}
where $\elab$ is the laboratory energy of the scattered electron.  The 
unpolarized cross section $\sigma_u(\elab)$ and longitudinal polarization 
asymmetry $A(s_z;\elab)$ are shown as functions of scattered electron energy 
in Fig.~\ref{fg:five}.  The order-$\alpha^2$ quantities are shown as dashed 
curves in parts~(a) and (b) of the figure.  The fractional 
correction to the unpolarized cross section $\sigma^{(1)}/\sigma^{(0)}$ is 
shown as the solid curve in part~(a) of the figure.  It increases from 
$-$0.2\% near the endpoint at 17.36~GeV to $+$0.2\% at the beam energy.  
The correction to the asymmetry function $\Delta A(s_z;\elab)$ is shown as the 
solid curve in part~(b) of the figure.  Near the endpoint, $\Delta A$ 
is almost exactly one thousand times smaller than the order-$\alpha^2$ 
asymmetry.  It becomes fractionally larger near the zero of $A^{(0)}$ at 
25.15~GeV.  

The effects of the order-$\alpha^3$ corrections upon the analyzing powers of 
the seven active channels of the Cherenkov detector are listed in 
Table~\ref{tab:one}.  The nominal acceptance in scattered energy, the 
order-$\alpha^2$ analyzing power $\apr_j^{(0)}$, and the order-$\alpha^3$ 
fractional correction to the analyzing power are listed for each channel.  The 
SLC beam polarization is determined from the channels near the endpoint (5-7).  
It is clear that proper inclusion of the radiative corrections increases the 
analyzing powers by 0.1\% of themselves.  This decreases the beam polarization 
by the same fractional amount.  Since the left-right asymmetry is the ratio of 
the measured $Z$-event asymmetry $A_Z$ and the beam polarization,
\begin{equation}
\alr =  \frac{A_Z}{\pole},
\end{equation}
the application of the order-$\alpha^3$ corrections {\it increases} the 
measured value of $\alr$ by 0.1\% of itself.  The corrections are much too 
small and have the wrong sign to account for the SLD/LEP discrepancy.

\subsection{The HERA Polarimeters}

The $e^\pm$ ring of the HERA $e^\pm$-$p$ collider is the first $e^\pm$ storage 
ring to operate routinely with polarized beams and it is the first storage ring 
to operate with a longitudinally polarized beam \cite{ref:HERAbeams}.  The ring 
is instrumented with transverse and longitudinal Compton polarimeters.  

\subsubsection{The Transverse Polarimeter}

The HERA transverse polarimeter \cite{ref:HERAtrans} collides 2.41~eV photons 
from a continuous-wave Argon-Ion laser with the 27.5~GeV HERA positron beam. 
 The scattered photons are separated from the electron beam by the dipole 
magnets of the accelerator lattice and are detected by a segmented 
tungsten-scintillator calorimeter located about 65~m from the $e^+$-$\gamma$ 
collision point.  The scattering rate is sufficiently small that the 
calorimeter measures the energy and vertical position of individual photons.  

When the positron beam is transversely polarized, the differential cross 
section depends upon the azimuthal directions of the scattered particles.  The 
average direction the scattered photons changes when the laser helicity is 
reversed.  The polarimeter measures the projected vertical direction $\theta_y$ 
and energy $\klab$ of each scattered photon.  The shift in the centroid of the 
$\theta_y$ distribution that occurs with helicity reversal 
$\delta\theta^{meas}_y(\klab)$ is proportional to the product of the photon 
polarization and the vertical positron polarization $\poley$,
\begin{equation}
\delta\theta^{meas}_y(\klab) = \langle\theta_y\rangle_- - 
\langle\theta_y\rangle_+ = \poley\cdot\polg\cdot\delta\theta_y(\klab),
\end{equation}
where $\delta\theta_y(\klab)$ is the shift for 100\% positron and photon 
polarizations.  This quantity is given by the following expression,
\begin{equation}
\delta\theta_y(\klab) = 2\frac{\int d\phi_\gamma^\prime 
\sigma_p(s_y;\klab,\phi_\gamma^\prime)\sin\theta_\gamma^\prime 
\sin\phi_\gamma^\prime} {\sigma_u(\klab)},
\end{equation}
where $\theta_\gamma^\prime$ and $\phi_\gamma^\prime$ are the polar angle and 
azimuth of the scattered photon in the laboratory frame.  Note that 
$\theta_\gamma^\prime$ is a constant for fixed $\klab$.    

\begin{figure}[t]
\centering
\epsfig{file=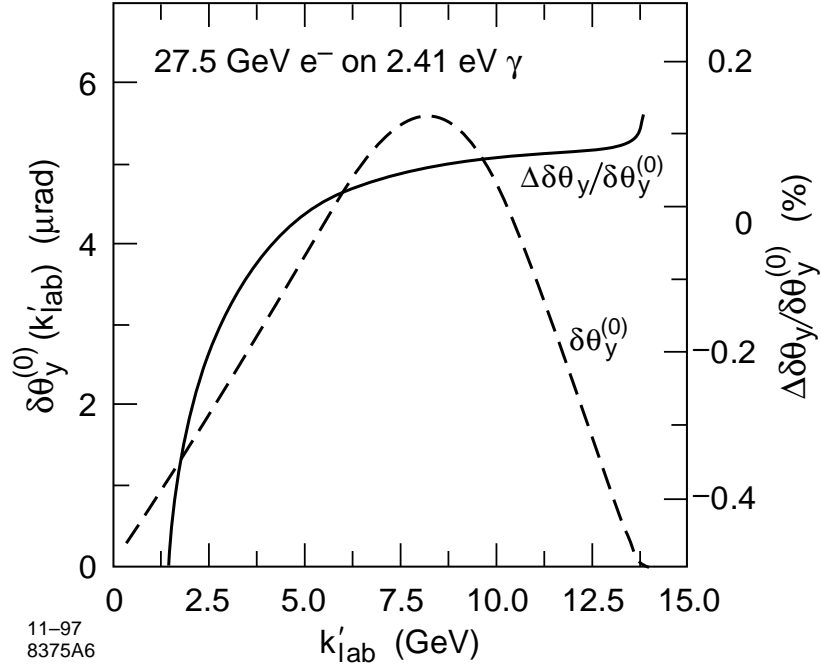,height=3.5in}
\caption{The order-$\alpha^2$ function $\delta\theta_y^{(0)}(\klab)$ is plotted 
as the dashed curve as a function of the laboratory energy of the scattered 
photon for the case of a 27.5~GeV positron colliding with a 2.41~eV photon.  
The fractional difference in the fully order-$\alpha^3$ corrected function and 
the order-$\alpha^2$ function is shown the solid curve.}
\label{fg:six}
\end{figure}

The order-$\alpha^3$ corrections modify the function $\delta\theta_y(\klab)$.  
The exact modification depends upon the details of how the polarimeter reacts 
to the two-photon final state.  It is assumed that the segmented calorimeter of 
the HERA transverse polarimeter cannot distinguish between one-photon and 
two-photon final states.  The energy measured by the calorimeter for two-photon 
final states is then the sum of the individual photon energies 
$\klab=\klab(1)+\klab(2)$.  The measured vertical angle is assumed to the the 
energy-weighted mean of the individual photon angles, $\theta_y = 
[\theta_y(1)\klab(1)+\theta_y(2)\klab(2)]/\klab$.  The order-$\alpha^2$ 
function $\delta\theta_y^{(0)}(\klab)$ is plotted as function of laboratory 
photon energy in Fig.~\ref{fg:six}.  The maximum angular separation of 
5.6~$\mu$m occurs near 8~GeV.  The fractional change caused by the 
order-$\alpha^3$ corrections $\Delta\delta\theta_y/\delta\theta_y^{(0)}$ is 
also shown as a function of $\klab$.  Note that the correction is typically 
$+$0.08\% near the maximum separation which would lower the measured transverse 
polarization by the same fractional amount.

\subsubsection{The Longitudinal Polarimeter}

A longitudinal polarimeter at HERA has been built by the HERMES Collaboration 
\cite{ref:HERAlong}.  The 27.5~GeV HERA positron beam is brought into collision 
with a 2.33~eV photon beam produced by a pulsed frequency-doubled Nd:YAG 
laser. The scattered photons are separated from the electron beam by the dipole 
magnets of the accelerator lattice and are detected by an array of NaBi 
crystals.  Since several thousand scattered photons are produced on each pulse, 
it is not possible to measure the cross section asymmetry as a function of 
photon energy.  Instead, the calorimeter measures the asymmetry in deposited 
energy $A_E$ as the photon helicity is reversed, 
\begin{equation}
A_E = \frac{E_-^{dep}-E_+^{dep}}{E_-^{dep}+E_+^{dep}} = \pole\polg\apr_E,
\end{equation}
where $E^{dep}_\pm$ is the energy deposited by all accepted photons in the 
crystal calorimeter.  The analyzing power $\apr_E$ is given by the following 
expression,
\begin{equation}
\apr_E = \frac{\int d\klab \klab R(\klab)\sigma_p(s_z;\klab)} {\int d\klab 
\klab R(\klab)\sigma_u(\klab)}
\end{equation}
where $R(\klab)$ describes the response of the detector.  For this estimate, it 
is assumed that the calorimeter has uniform response in energy from the minimum 
accepted energy of 56~MeV (lower energy photons miss the calorimeter) to the 
maximum energy of 13.62~GeV.  The order-$\alpha^2$ analyzing power and the full 
order-$\alpha^3$ correction are
\begin{eqnarray}
\apr_E^{(0)} &=& 0.1838 \\
\frac{\apr_E-\apr_E^{(0)}} {\apr_E^{(0)}} &=& +0.20\%.
\end{eqnarray}
The fractional correction to the longitudinal polarization scale is therefore 
$-$0.20\%.

\subsection{A Linear Collider Polarimeter}

\begin{figure}[t]
\centering
\epsfig{file=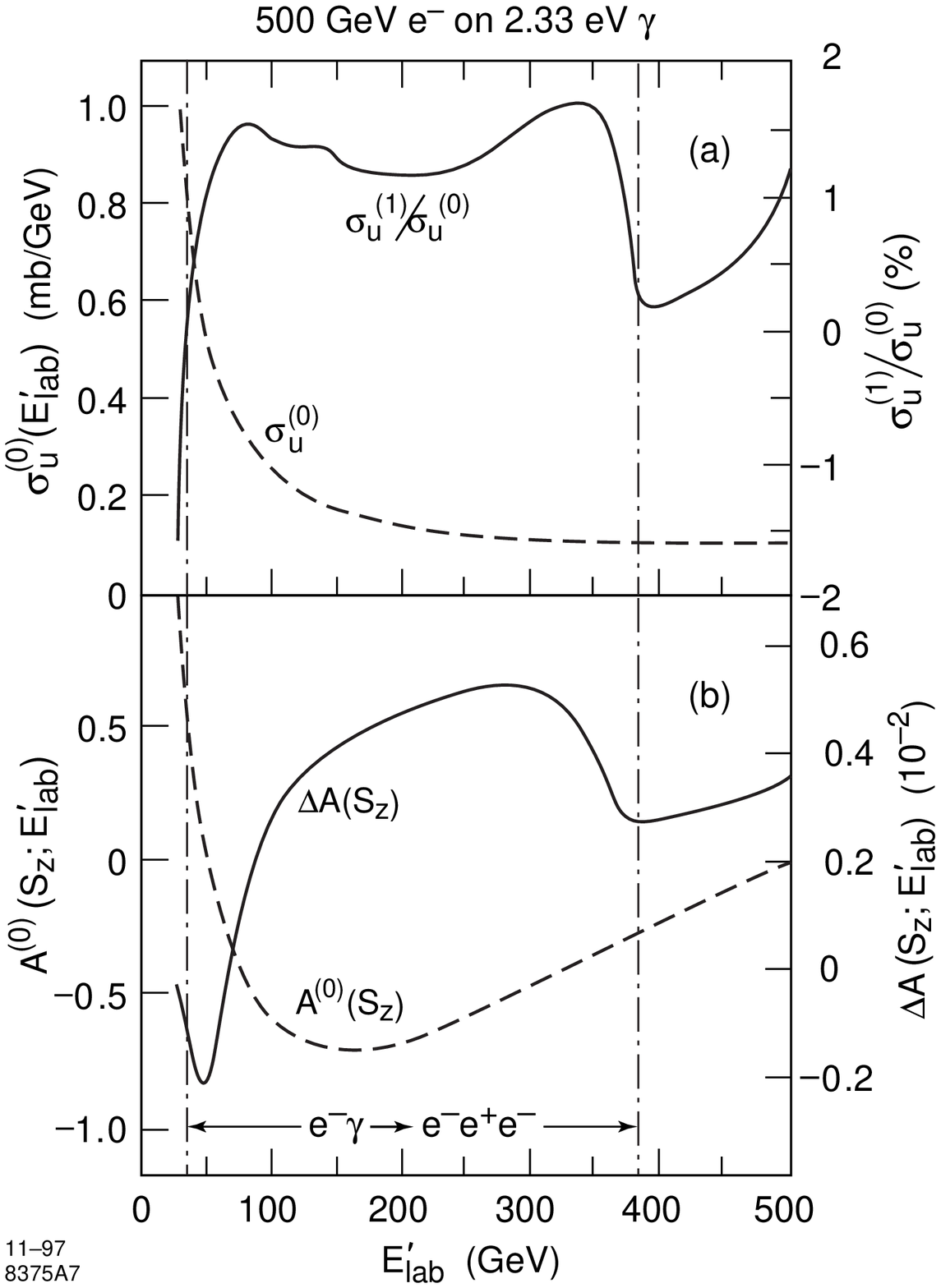,height=7in}
\caption{The order-$\alpha^2$ cross section $\sigma^{(0)}_u$ and the ratio 
$\sigma^{(1)}_u/\sigma^{(0)}_u$ are shown in part~(a) as functions of the 
laboratory energy of the scattered electron for the case of a 500~GeV electron 
colliding with a 2.33~eV photon.  The order-$\alpha^2$ asymmetry function and 
the difference in the fully order-$\alpha^3$ corrected asymmetry and the 
order-$\alpha^2$ asymmetry are shown as functions of the laboratory energy of 
the scattered electron in part~(b).  The vertical dashed-dotted lines 
indicate the allowed kinematic region for electrons from the $e^-\gamma\to 
e^-e^+e^-$ subprocess.}
\label{fg:seven}
\end{figure}

Longitudinally polarized beams are likely to be important features of a future 
Linear Collider.  It is assumed that any such machine will include SLC-like 
polarimetry which detects and momentum-analyzes scattered electrons.  The 
unpolarized cross section $\sigma_u(\elab)$ and longitudinal polarization 
asymmetry $A(s_z;\elab)$ are shown as functions of scattered electron energy in 
Fig.~\ref{fg:seven} for the case of a 500~GeV electron beam colliding with a 
2.33~eV photon beam.  The order-$\alpha^2$ quantities are shown as dashed 
curves in parts~(a) and (b) of the figure.  The cross section is 
largest near the backscattering edge at 26.42~GeV.  The longitudinal asymmetry 
function is 0.9944 at the kinematic endpoint.  It decreases rapidly with 
increasing energy and passes through zero at 50.19~GeV.  The fractional 
correction to the unpolarized cross section $\sigma^{(1)}/\sigma^{(0)}$ is 
shown as the solid curve in part~(a) of the figure.  It increases from 
$-$1.6\% near the endpoint to $+$1.2\% at the beam energy.  Superimposed upon 
this is the contribution of the $e^-e^+e^-$ final state which is kinematically 
constrained to the region $34.36~{\rm GeV}<\elab<386.1~{\rm GeV}$.  The effect 
of this final state is to increase the correction to the 1.0-1.7\% level in the 
kinematically allowed region.  The correction to the asymmetry function $\Delta 
A(s_z;\elab)$ is shown as the solid curve in part~(b) of the figure.  
Near the endpoint, $\Delta A$ is $-$4$\times$10$^{-4}$ and represents a 
negligible correction.   Due to the influence of the $e^-e^+e^-$ final state, 
it decreases to $-$2.2$\times$10$^{-3}$ near 49~GeV then begins to increase to 
$+$5.3$\times$10$^{-3}$ near 306~GeV where it is a 1\% correction to the 
asymmetry function.   

\section{Summary}
\label{sec:summary}

The construction of a computer code, COMRAD, to calculate the cross sections 
for the spin-polarized processes $e^-\gamma\to 
e^-\gamma,e^-\gamma\gamma,e^-e^+e^-$ to order-$\alpha^3$ has been described.  
The code is based upon the work of Tsai, DeRaad, and Milton \cite{ref:tdrm} for 
the virtual and soft-photon corrections.  The hard-photon photon corrections 
and the application of the virtual corrections to arbitrary electron spin 
direction are based upon the work of G\'ongora and Stuart \cite{ref:gs}.  The 
calculation of the cross section for the $e^-e^+e^-$ final state was performed 
by the author.  As implemented, the code calculates cross sections for 
circularly-polarized initial-state photons and arbitrarily polarized 
initial-state electrons.  Final-state polarization information is not presented 
to a user of the code but is present at the matrix element level.  The 
modification of the code to extract this information would not be difficult.  

The order-$\alpha^3$ corrections to the longitudinal polarization asymmetry 
calculated by COMRAD agree well with those of Veltman \cite{ref:veltman}.  
However, the order-$\alpha^3$ corrections to the unpolarized cross section 
calculated by COMRAD do not agree with those of Veltman.  

The application of the code to the SLD Compton polarimeter indicates that the 
order-$\alpha^3$ corrections produce a fractional shift in the SLC polarization 
scale of $-$0.1\%.  This shift is much too small and of the wrong sign to 
account for the discrepancy in the Z-pole asymmetries measured by the SLD 
Collaboration and the LEP Collaborations. 

The application of the code to the photon-based polarimeters at the HERA 
storage ring indicates that the order-$\alpha^3$ corrections also have small 
effect on the measurements of the HERA positron polarization.  The effects on 
the transverse polarization measurements are typically less than 0.1\%.  The 
effect upon the calibration of the HERMES longitudinal polarimeter is a 
somewhat larger 0.2\%.

The application of the code to a polarimeter at a future Linear Collider 
indicates that the order-$\alpha^3$ corrections are very small near 
the Compton edge but increase to the 1\% level elsewhere.  The $e^-e^+e^-$ 
final state contributes significantly to the net corrections.

\acknowledgments

This work was supported by Department of Energy contract DE-AC03-76SF00515.  
The author would like to thank Robin Stuart and B.F.L.~Ward for their helpful 
comments on this manuscript.  Bennie also pointed out an additional discrepancy between the preprint and published versions of Ref.~\cite{ref:gs}.

\appendix

\section{Errata}
\label{sec:errata}

The following typographical errors were found in Ref.~\cite{ref:gs}:
\begin{enumerate}
\item
All of the spinor products of the form $\bar{u}_\pm(q_1)\!{\not\!p}u_\mp(q_2)$ 
given in Eqs.~3.3-3.10 and C.1-C.8 formally vanish (they are the traces of an 
odd number of gamma matrices) and should be replaced by 
$\bar{u}_\pm(q_1)\!{\not\!p}u_\pm(q_2)$ (the helicities of the 
$\bar{u}$-spinors are correct in all cases and the helicities of the 
$u$-spinors are wrong in all cases).  The right-hand-sides of the spinor 
product definitions are nearly all correct (see items~4 and 5 below).
\item
The sign of the second term in square brackets on the right-hand-side of 
Eq.~3.5.1 in the preprint is incorrect 
[$-\spl(\pt,q^{\prime\prime})\cdots$ should be 
$+\spl(\pt,q^{\prime\prime})\cdots$].  Unfortunately, a serious typesetting 
error in the published version significantly altered the equation.  The correct 
equation should read as follows,
\begin{eqnarray}
D_{++-}(q,\pt;q^\prime,\pt;q^{\prime\prime},\po) & = &
-\frac{\spl(\po,\pt)\smn(\po,q)}
{\spl(\pt,q)\spl(\pt,q^\prime)\smn(\po,q^{\prime\prime})}\ \ \ \ \ \ \ \ \ \ 
\ \ \ \nonumber \\
&\times&\biggl[\frac{\spl(\pop,q^{\prime\prime})\smn(\pop,\ptp)}{m^2}
\bar{u}_-(\po)\!{\not\!p}_bu_-(\pt)\bar{u}_-(q^\prime)\!{\not\!p}_au_-(\pt) 
\nonumber \\
&&+\spl(\pt,q^{\prime\prime})\smn(\po,\ptp)
\bar{u}_-(q^\prime)\!{\not\!p}_au_-(\pt) \biggr]. \nonumber
\end{eqnarray}
\item
The sign of the fourth term in square brackets on the right-hand-side of 
Eq.~3.6.1 is incorrect [$-\spl(\pt,q^\prime)\cdots$ should be 
$+\spl(\pt,q^\prime)\cdots$].
\item
The right-hand-side of the first of Eqs.~3.6.2 is incorrect.  The quantities 
$p_1$ and $p_2$ should be replaced by $\pop$ and $\ptp$, respectively.
\item
The left-hand-side of the second of Eqs.~3.7.2 is incorrect.  The quantity 
$\pop$ should be replaced by $\pt$.  The right-hand-side is also incorrect.  
The sign of the second term should be flipped 
[$-\smn(q^{\prime\prime},\ptp)\cdots$ should be 
$+\smn(q^{\prime\prime},\ptp)\cdots$].
\item
The second factor in the second term in square brackets on the right-hand-side 
of Eq.~3.8.1 should be $\smn(\po,q^{\prime\prime})$ instead of 
$\smn(\pop,q^{\prime\prime})$.
\item
The fourth factor in the fourth term in square brackets on the right-hand-side 
of Eq.~C.7.1 should be $\smn(\ptp,q^\prime)$ instead of $\smn(\ptp,q)$.
\item
The heading of Eqs.~4.10 which states that they define quantities of the form 
$\epsilon_-^\prime{\cal L}_i\epsilon_+$ is correct and all of the 
left-hand-sides which state the reverse helicity configuration are wrong.
\item
The heading of Eqs.~4.11 which states that they define quantities of the form 
$\epsilon_+^\prime{\cal L}_i\epsilon_-$ is correct and all of the 
left-hand-sides which state the reverse helicity configuration are wrong.
\end{enumerate}

The following typographical errors were found in Ref.~\cite{ref:tdrm}:
\begin{enumerate}
\item
The signs of two of the tree-level helicity amplitudes given in Eqs.~5 are 
incorrect.  The signs of the amplitudes $f^{(2)}(-+;+-)$ [the third amplitude] 
and $f^{(2)}(-+;++)$ [the fifth amplitude] should be reversed.
\item
The sign of the order-$\alpha^2$ amplitude $f^{(4)}(-+;+-)$ defined in Eq.~8 
should also be reversed.
\end{enumerate}

\section{The Matrix Element for $e^-\gamma\to e^-e^+e^-$}
\label{sec:mateee}

\begin{figure}[t]
\centering
\epsfig{file=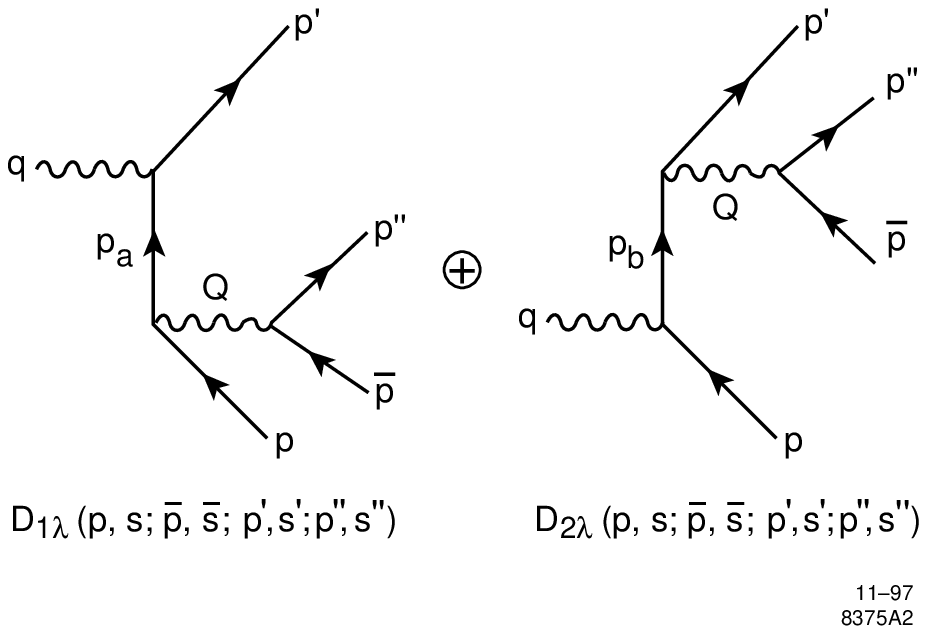,height=2.5in}
\caption{Two amplitudes for the process $\gamma\to e^-e^+e^-e^+$ (time flows 
left to right).}
\label{fg:two}
\end{figure}

The matrix for the process $e^-(s)\gamma(\lambda)\to 
e^-(s^\prime)e^+(\bar{s})e^-(s^{\prime\prime})$ is calculated from the two 
amplitudes for the process $\gamma(\lambda)\to e^+(s) 
e^-(s^\prime)e^+(\bar{s})e^-(s^{\prime\prime})$ shown in Fig.~\ref{fg:two}.  
This formulation is chosen so that each term in Eq.~\ref{eq:meee} has exactly 
one negative momentum.  The internal momenta shown in the Fig.~\ref{fg:two} are 
defined as,
\begin{eqnarray}
Q & = & p^{\prime\prime} + \bar{p} \\
p_a & = & -\left(p+Q\right) = p^\prime - q \\
p_b & = & q - p = p^\prime + Q.
\end{eqnarray}

Using the techniques described in Ref.~\cite{ref:gs} and the Chisholm identity 
\cite{ref:ks},
\begin{equation}
\gamma^\mu\bar{u}_\pm(q_1)\gamma_\mu u_\pm(q_2) = 2 \left[ 
u_\pm(q_2)\bar{u}_\pm(q_1) + u_\mp(q_1)\bar{u}_\mp(q_2) \right],
\end{equation}
it is straightforward to evaluate the amplitudes $D_{1\lambda}$ and 
$D_{2\lambda}$.  Unfortunately, the exact form for each of these functions 
depends upon the initial-state photon helicity $\lambda$.  They are listed 
below:
\begin{eqnarray}
\lefteqn{D_{1+}(p,s;\bar{p},\bar{s};p^\prime,s^\prime;p^{\prime\prime},
s^{\prime\prime}) = \frac{2\sqrt{2}e^3}{-2p^\prime\cdot q\ 
Q^2\spl(\hatq,q)}\Biggl\{ } \nonumber \\ &\ \ &
\smn(\ptp,q)\spl(\po,\pt)\spl(\hatq,\pob)\smn(\ptpp,\pt) + 
\spl(\hatq,\pop)\smn(\pop,\ptp)\smn(q,\ptpp)\spl(\pob,\po) \nonumber \\
&\ \ & + \smn(\ptp,q)\spl(\pob,\po)\Bigl[\spl(\hatq,\pop)\smn(\pop,\ptpp) + 
\spl(\hatq,\ptp)\smn(\ptp,\ptpp) - \spl(\hatq,q)\smn(q,\ptpp) \Bigr]
\nonumber \\ &\ \ & 
+\frac{\spl(\hatq,\pop)\smn(\pop,\ptp)\spl(\po,\pt)\smn(\ptpp,\pt)}{m^2}\Bigl[ 
\smn(q,\pop)\spl(\pop,\pob) + \smn(q,\ptp)\spl(\ptp,\pob) \Bigr] \nonumber \\ 
&\ \ & 
+ \frac{\smn(\popp,\ptpp)\spl(\ptb,\pob)}{m^2} \biggl( \nonumber \\ &\ \ & 
\smn(\ptp,q)\spl(\po,\pt)\spl(\hatq,\popp)\smn(\ptb,\pt) + 
\spl(\hatq,\pop)\smn(\pop,\ptp)\smn(q,\ptb)\spl(\popp,\po) \nonumber \\
&\ \ & + \smn(\ptp,q)\spl(\popp,\po)\Bigl[\spl(\hatq,\pop)\smn(\pop,\ptb) + 
\spl(\hatq,\ptp)\smn(\ptp,\ptb) - \spl(\hatq,q)\smn(q,\ptb) \Bigr]  \nonumber 
\\ &\ \ & 
+\frac{\spl(\hatq,\pop)\smn(\pop,\ptp)\spl(\po,\pt)\smn(\ptb,\pt)}{m^2}\Bigl[ 
\smn(q,\pop)\spl(\pop,\popp) + \smn(q,\ptp)\spl(\ptp,\popp) \Bigr] \biggr) 
\Biggr\} \label{eq:dop} \\
& & \nonumber \\
\lefteqn{D_{1-}(p,s;\bar{p},\bar{s};p^\prime,s^\prime;p^{\prime\prime},
s^{\prime\prime}) = \frac{2\sqrt{2}e^3}{-2p^\prime\cdot q\ 
Q^2\smn(q,\hatq)}\Biggl\{ } \nonumber \\ &\ \ &
\smn(\ptp,\hatq)\spl(\po,\pt)\spl(q,\pob)\smn(\ptpp,\pt) + 
\spl(q,\pop)\smn(\pop,\ptp)\smn(\hatq,\ptpp)\spl(\pob,\po) \nonumber \\
&\ \ & + \smn(\ptp,\hatq)\spl(\pob,\po)\Bigl[\spl(q,\pop)\smn(\pop,\ptpp) + 
\spl(q,\ptp)\smn(\ptp,\ptpp) \Bigr] \nonumber \\
&\ \ & + \frac{\spl(q,\pop)\smn(\pop,\ptp)\spl(\po,\pt)\smn(\ptpp,\pt)}{m^2}\Bigl[ 
\smn(\hatq,\pop)\spl(\pop,\pob) + \smn(\hatq,\ptp)\spl(\ptp,\pob) \nonumber \\ 
&\ \ & -  \smn(\hatq,q)\spl(q,\pob) \Bigr]
 + \frac{\smn(\popp,\ptpp)\spl(\ptb,\pob)}{m^2} \biggl( \nonumber \\ &\ \ & 
\smn(\ptp,\hatq)\spl(\po,\pt)\spl(q,\popp)\smn(\ptb,\pt) + 
\spl(q,\pop)\smn(\pop,\ptp)\smn(\hatq,\ptb)\spl(\popp,\po) \nonumber \\
&\ \ & + \smn(\ptp,\hatq)\spl(\popp,\po)\Bigl[\spl(q,\pop)\smn(\pop,\ptb) + 
\spl(q,\ptp)\smn(\ptp,\ptb)  \Bigr] \nonumber \\
&\ \ &  +\frac{\spl(q,\pop)\smn(\pop,\ptp)\spl(\po,\pt)\smn(\ptb,\pt)}{m^2} 
\Bigl[ \smn(\hatq,\pop)\spl(\pop,\popp)  \nonumber \\
&\ \ & + \smn(\hatq,\ptp)\spl(\ptp,\popp) - \smn(\hatq,q)\spl(q,\popp) \Bigr] 
\biggr) \Biggr\}  \label{eq:dom} \\
& & \nonumber \\
\lefteqn{D_{2+}(p,s;\bar{p},\bar{s};p^\prime,s^\prime;p^{\prime\prime},
s^{\prime\prime}) = \frac{2\sqrt{2}e^3}{-2p\cdot q\ Q^2\spl(\hatq,q)}
\Biggl\{ } \nonumber \\ &\ \ &
- \smn(\pop,\ptp)\spl(\hatq,\po)\spl(\pop,\pob)\smn(\ptpp,q) + 
\spl(\po,\pt)\smn(q,\pt)\smn(\ptp,\ptpp)\spl(\pob,\hatq) \nonumber \\
&\ \ & - \spl(\hatq,\po)\smn(\ptp,\ptpp)\Bigl[\spl(\pob,\po)\smn(\po,q) + 
\spl(\pob,\pt)\smn(\pt,q) \Bigr] \nonumber \\
&\ \ & - \frac{\spl(\pop,\pob)\smn(\pop,\ptp)\spl(\po,\pt)\smn(q,\pt)}{m^2}\Bigl[ 
\smn(\ptpp,q)\spl(q,\hatq) - \smn(\ptpp,\po)\spl(\po,\hatq) \nonumber \\
 &\ \ & -  \smn(\ptpp,\pt)\spl(\pt,\hatq) \Bigr]
 + \frac{\smn(\popp,\ptpp)\spl(\ptb,\pob)}{m^2} \biggl( \nonumber \\ 
&\ \ & - \smn(\pop,\ptp)\spl(\hatq,\po)\spl(\pop,\popp)\smn(\ptb,q) + 
\spl(\po,\pt)\smn(q,\pt)\smn(\ptp,\ptb)\spl(\popp,\hatq) \nonumber \\
&\ \ & - \smn(\ptp,\ptb)\spl(\hatq,\po)\Bigl[\spl(\popp,\po)\smn(\po,q) + 
\spl(\popp,\pt)\smn(\pt,q)  \Bigr] \nonumber \\
&\ \ &  -\frac{\spl(\pop,\popp)\smn(\pop,\ptp)\spl(\po,\pt)\smn(q,\pt)}{m^2} 
\Bigl[ \smn(\ptb,q)\spl(q,\hatq)  \nonumber \\
&\ \ & - \smn(\ptb,\po)\spl(\po,\hatq) - \smn(\ptb,\pt)\spl(\pt,\hatq) \Bigr] 
\biggr) \Biggr\}   \label{eq:dtp} \\
& & \nonumber \\
\lefteqn{D_{2-}(p,s;\bar{p},\bar{s};p^\prime,s^\prime;p^{\prime\prime},
s^{\prime\prime}) = \frac{2\sqrt{2}e^3}{-2p\cdot q\ Q^2\smn(q,\hatq)}
\Biggl\{ } \nonumber \\ &\ \ &
- \smn(\pop,\ptp)\spl(q,\po)\spl(\pop,\pob)\smn(\ptpp,\hatq) + 
\spl(\po,\pt)\smn(\hatq,\pt)\smn(\ptp,\ptpp)\spl(\pob,q) \nonumber \\
&\ \ & + \spl(\hatq,\po)\smn(\ptp,\ptpp)\Bigl[\spl(\pob,q)\smn(q,\hatq) - 
\spl(\pob,\po)\smn(\po,\hatq) - \spl(\pob,\pt)\smn(\pt,\hatq)\Bigr] \nonumber 
\\ &\ \ & 
+ \frac{\spl(\pop,\pob)\smn(\pop,\ptp)\spl(\po,\pt)\smn(\hatq,\pt)}{m^2}\Bigl[ 
\smn(\ptpp,\po)\spl(\po,q) + \smn(\ptpp,\pt)\spl(\pt,q) \Bigr] 
\nonumber \\ &\ \ & 
 + \frac{\smn(\popp,\ptpp)\spl(\ptb,\pob)}{m^2} \biggl( \nonumber \\ 
&\ \ & - \smn(\pop,\ptp)\spl(q,\po)\spl(\pop,\popp)\smn(\ptb,\hatq) + 
\spl(\po,\pt)\smn(\hatq,\pt)\smn(\ptp,\ptb)\spl(\popp,q) \nonumber \\
&\ \ & + \smn(\ptp,\ptb)\spl(q,\po)\Bigl[\spl(\popp,q)\smn(q,\hatq) - 
\spl(\popp,\po)\smn(\po,\hatq)-  \spl(\popp,\pt)\smn(\pt,\hatq)  \Bigr] 
\nonumber \\ 
&\ \ & +\frac{\spl(\pop,\popp)\smn(\pop,\ptp)\spl(\po,\pt)\smn(\hatq,\pt)}
{m^2} \Bigl[ \smn(\ptb,\po)\spl(\po,q) + \smn(\ptb,\pt)\spl(\pt,q) \Bigr] 
\biggr) \Biggr\}
\label{eq:dtm}
\end{eqnarray}

\section{Operational Details}
\label{sec:opdet}

The Fortran-code COMRAD \cite{ref:fortran} consists of a main program COMRAD 
that controls the three weighted Monte Carlo generators: COMTN2, COMEGG, and 
COMEEE.  These perform integrations of the cross sections for the $e^-\gamma$, 
$e^-\gamma\gamma$, and $e^-e^+e^-$ final states, respectively.  All three 
generators use a common set of conventions, input parameters, and a common 
interface routine called WGTHST.  The routine WGTHST permits the user to 
accumulate event weights in a manner that is appropriate to his/her needs. 
Note that all quantities discussed in this section are assumed to be of type 
REAL*8 unless otherwise specified. 

\subsection{The Program COMRAD}

The program COMRAD initializes all quantities and sequentially calls each of 
the event generators.  Communication with the generators occurs through the 
/CONTROL/ common block which is specified within COMRAD.  This common block 
contains the variables: EB, EPHOT, XME, XMG, KGMIN, ALPHA, PI, ROOT2, BARN, 
SPIN(3), LDIAG, LBF, and NTRY.

The variables EB and EPHOT specify the energy and laboratory frame to be used 
in the calculation.  It is assumed that the incident electron is moving in the 
$+$z-direction with an energy EB GeV (${\rm EB}\geq m$).  The incident photon 
is assumed to be moving in the $-$z-direction with an energy EPHOT GeV.
The spin of the initial-state electron is specified in its rest frame by the 
three-vector SPIN(3).  The maximum energy of additional soft-photons and the 
minimum energy of hard-photons is also specified in the electron rest frame 
($k_\gamma^{min}$) by the variable KGMIN.  The integer variable NTRY sets the 
number of trials for each of the event generators (COMTN2 and COMEGG generate 
NTRY trials whereas COMEEE produces smaller event weights and generates NTRY/20 
trials).  The logical flags LDIAG and LBF activate the calculation of the 
Tsai-DeRaad-Milton and Brown-Feynman expressions for the corrections to the 
unpolarized cross section in COMTN2 (for diagnostic purposes).  The common 
block /CONTROL/ also contains several constants used by the generators.

\subsection{The Generator COMTN2}

The subroutine COMTN2 simulates the two-body $e^-\gamma$ final states.  The 
calculation is carried out in the center-of-mass frame using Eq.~\ref{eq:jdef} 
and Eqs.~\ref{eq:wgti}-\ref{eq:wgtiv} to calculate the event weights 
$W_1$-$W_4$.  The density function $\rho^{(2)}(\Omega_e^\prime)$ is chosen to 
be uniform in the polar variables $\cos\theta_e^\prime$ and $\phi_e^\prime$,
\begin{equation} 
\rho^{(2)}(\Omega_e^\prime)=\frac{N_{trial}}{4\pi},
\end{equation}
where $N_{trial}$ is the number of event trials.

\subsection{The Generator COMEGG}

The subroutine COMEGG simulates the three-body $e^-\gamma\gamma$ final states.  
The calculation is carried out in the center-of-mass frame using 
Eq.~\ref{eq:eggxs} and Eqs.~\ref{eq:wgtiii} and \ref{eq:wgtiv} to calculate the 
event weights $W_3$ and $W_4$ ($W_1$ and $W_2$ are always returned as 0).  The 
five quantities $E_e^\prime$, $\cos\theta_e^\prime$, $\phi_e^\prime$, 
$E_\gamma^\prime$, and $\phi_\gamma^\prime$ are chosen according to the density 
function $\rho^{(5)}$ as follows,
\begin{equation}
\rho^{(5)}(E_e^\prime,\Omega_e^\prime,E_\gamma^\prime \phi_\gamma^\prime) = 
\frac{N_{trial}}{4\pi\cdot2\pi}\cdot\frac{C_e}{E_e^{max}+E_\gamma^{min}
-E_e^\prime}\cdot C_\gamma\left[\frac{1}{E_\gamma^\prime}
+\frac{1}{E_\gamma^{max}+E_\gamma^{min}-E_\gamma^\prime}\right],
\end{equation}
where: $N_{trial}$ is the number of generated trials (some are later 
discarded); $E_e^{max}$ and $E_\gamma^{max}$ are the maximum electron and 
photon energies in the cm-frame; $E_\gamma^{min}$ is the minimum photon energy 
in the cm-frame; and $C_e$ and $C_\gamma$ are normalization constants given as 
follows,
\begin{eqnarray}
C_e & = & \frac{1}{\ln\left[(E_e^{max}+E_\gamma^{min}-m)
/E_\gamma^{min}\right]} \\
C_\gamma & = & \frac{1}{2\ln\left[E_\gamma^{max}/E_\gamma^{min}\right]}. 
\end{eqnarray}
The minimum energy in the cm-frame is related to the minimum photon energy in 
the initial-state electron rest frame as follows,
\begin{equation}
E_\gamma^{min} = \frac{m}{E_{cm}}k_\gamma^{min},
\end{equation} 
where $E_{cm}$ is the total center-of-mass energy.  Note that a photon emitted 
in the cm-frame in the $-z$ direction with energy $E_\gamma^{min}$ has an 
energy $k_\gamma^{min}$ in the electron rest frame.  If emitted in any other 
direction, it has a smaller energy in the electron rest frame.

After all five variables have been chosen, the electron and photon energies are 
checked for consistency with three-body kinematics (the angle 
$\theta_{e\gamma}$ between the electron and photon directions must satisfy the 
condition $|\cos\theta_{e\gamma}|\leq 1$).  If this condition is not 
satisfied, the trial is discarded.  If it is satisfied, the four-vectors 
$p^\prime$, $q^\prime$, and $q^{\prime\prime}$ are generated.  The photon 
energies in the initial-state electron rest frame are then calculated and if 
either is found 
to be less than $k_\gamma^{min}$, the trial is discarded.  The kinematical 
boundary of the integration is therefore exactly the same as the one that 
defines the upper-limit of the soft-photon integration (and the function $J$).  
The integrated region is thus the complement of the soft-photon region and the 
sum of the cross sections returned by COMTN2 and COMEEG is independent of 
$k_\gamma^{min}$.  The event generation procedure retains approximately 60\% of 
the generated trials over a wide range incident electron and photon energies.

\subsection{The Generator COMEEE}

The subroutine COMEEE simulates the three-body $e^-e^+e^-$ final states.  The 
calculation is carried out in the center-of-mass frame using Eq.~\ref{eq:eeexs} 
and Eqs.~\ref{eq:wgtiii} and \ref{eq:wgtiv} to calculate the event weights 
$W_3$ and $W_4$ ($W_1$ and $W_2$ are always returned as 0).  The five 
quantities $E_e^\prime$, $\cos\theta_e^\prime$, $\phi_e^\prime$, 
$E_e^{\prime\prime}$, and $\phi_e^{\prime\prime}$ are chosen according to the 
density function $\rho^{(5)}$ as follows,
\begin{equation}
\rho^{(5)}(E_e^\prime,\Omega_e^\prime,E_e^{\prime\prime},\phi_e^{\prime\prime}) 
= \frac{N_{trial}}{4\pi\cdot2\pi\cdot (E_e^{max}-m)^2},
\end{equation}
where $N_{trial}$ is the number of generated trials (some are later discarded) 
and $E_e^{max}$ is the maximum electron energy in the cm-frame.  Note that the 
use of uniform phase space works well near threshold (the polarimetry case) but 
is inadequate at very high energies.

After all five variables have been chosen, the electron energies are checked 
for consistency with three-body kinematics (the angle 
$\theta_{e^\prime e^{\prime\prime}}$ between the electron directions must 
satisfy the condition $|\cos\theta_{e^\prime e^{\prime\prime}}|\leq 1$).  If 
this condition is not satisfied, the trial is discarded.  This procedure 
retains approximately 70\% of the generated trials near the $e^-e^+e^-$ 
threshold.

\subsection{The Interface Routine WGTHST}

The routine WGTHST allows the user to accumulate the information needed for 
his/her purposes.  The routine is called by the main program once before any 
event generation to permit initialization.  It is called by each of the event 
generators (COMTN2, COMEGG, and COMEEE) at the end of each event trial.  And 
finally, it is called by the main program after return from the last generator 
to permit the information to be output.

All communication with the routine occurs through the argument list,
\begin{center}
SUBROUTINE~WGTHST(IFLAG,NEM,PP,NGAM,QP,NEP,PB,WGT),
\end{center}
where: IFLAG is an integer flag which indicates the initialization call (0), an 
accumulation call (1), or the output call (2); NEM is an integer which 
indicates the number of electrons in the final state (1 or 2); PP(4,2) contains 
the four-vectors of the NEM electrons in the laboratory frame; NGAM is an 
integer which indicates the number of photons in the final state (0-2); 
QP(4,2) contains the four-vectors of the NGAM photons in the laboratory frame; 
NEP is an integer which indicates the number of positrons in the final state 
(0 or 1); PB(4) is the four-vector of the positron; and WGT(4) contains the 
four weights $W_1$-$W_4$ defined in Eqs.~\ref{eq:wgti}-\ref{eq:wgtiv}.  Note 
that the event weights have been defined such that correctly normalized total 
cross sections are obtained by summing the weights exactly once per call to 
WGTHST.  The calculation of final-state particle yields therefore requires that 
the weights be accumulated each time the given type of particle is encountered. 

% now the references. delete or change fake bibitem. delete next thre3
%   lines and directly read in your .bbl file if you use bibtex.

% \begin{table}
% \caption{}
% \label{}
% \end{tabular}
% \end{table}
\begin{table}
\caption{The effect of order$-\alpha^3$ radiative corrections upon the 
analyzing powers of the SLD Compton polarimeter.}
\label{tab:one}
\begin{tabular}{cccc} \hline
Channel & $E_e$ Acceptance & $\apr^{(0)}$ & $(\apr-\apr^{(0)})/\apr^{(0)}$ 
(\%) \\
\hline
7  & 17.14-18.02~GeV & 0.7133 & 0.096 \\
6  & 18.02-19.00~GeV & 0.6483 & 0.097 \\
5  & 19.00-20.11~GeV & 0.5520 & 0.103 \\
4  & 20.11-21.38~GeV & 0.4309 & 0.118 \\
3  & 21.38-22.83~GeV & 0.2851 & 0.153 \\
2  & 22.83-24.53~GeV & 0.1228 & 0.285 \\
1  & 24.53-26.51~GeV & $-$.0396 & $-$.673 \\
\end{tabular}
\end{table}


\begin{references}
\bibitem{ref:normnote} This expression differs by a factor of $2m$ from Eq.~1 
of Ref.~\cite{ref:tdrm} because the spinors in that work are normalized to 
unity.
\bibitem{ref:phinote} The azimuth $\phi^\prime$ is defined in any frame which 
has been rotated such that the electron direction defines the $+$z-axis.
\bibitem{ref:fortran} The code COMRAD generally conforms to the Fortran-77 
specification except that it performs COMPLEX*16 operations which are 
implemented in many compilers.  The code has been successfully compiled and run 
with the IBM-AIX, SUN-Solaris, and DEC-OpenVMS operating systems.
\end{references}
\end{document}